\documentclass[10pt]{article}
\usepackage{graphicx}
\usepackage{amsmath}
\usepackage{amssymb}
\usepackage{caption2}
\begin{document}
\bibliographystyle{prsty}
\begin{center}
{\large {\bf \sc{  Analysis of the doubly heavy  baryons in the nuclear matter  with   the   QCD sum rules  }}} \\[2mm]
Zhi-Gang Wang \footnote{E-mail,wangzgyiti@yahoo.com.cn.  }   \\
 Department of Physics, North China Electric Power University, Baoding 071003, P. R. China
\end{center}

\begin{abstract}
In this article, we  study the doubly heavy baryon states $\Xi_{cc}$, $\Omega_{cc}$,  $\Xi_{bb}$  and $\Omega_{bb}$
 in the nuclear matter using
the QCD sum rules, and  derive three coupled QCD sum rules for  the   masses, vector self-energies
and pole residues. The predictions for   the mass-shifts in the nuclear matter
$\Delta M_{\Xi_{cc}}=-1.11\,\rm{GeV}$, $\Delta M_{\Omega_{cc}}=-0.33\,\rm{GeV}$,
$\Delta M_{\Xi_{bb}}=-3.37\,\rm{GeV}$ and  $\Delta M_{\Omega_{bb}}=-1.05\,\rm{GeV}$
can be confronted with the experimental data in the future.
\end{abstract}

PACS numbers:  12.38.Lg; 14.20.Lq; 14.20.Mr

{\bf{Key Words:}}  Nuclear matter,  QCD sum rules

\section{Introduction}

In 2002,  the SELEX collaboration reported   the first observation
of a signal for the doubly charmed baryon state  $ \Xi_{cc}^+$ in
the charged decay mode $\Xi_{cc}^+\rightarrow\Lambda_c^+K^-\pi^+$ in the charm hadro-production experiment (E781) at Fermilab
\cite{SELEX2002}, and confirmed later by the same collaboration in
the decay mode $\Xi_{cc}^+\rightarrow pD^+K^- $ with the measured
mass $M_{\Xi_{cc}}=(3518.9 \pm 0.9) \,\rm{ MeV }$ \cite{SELEX2004}. And there have been several theoretical
approaches to calculate  the doubly heavy baryon masses  \cite{HH-B}.
In Ref.\cite{WangzgEPJA},  we study the $J^P={1\over 2}^+$ doubly heavy baryon states $\Omega_{QQ}$ and $\Xi_{QQ}$ by subtracting the contributions from the corresponding ${1\over 2}^-$ doubly heavy baryon states with the QCD sum rules, and make reasonable predictions for their masses.

 The  QCD sum rules are a powerful theoretical tool in studying the
ground state hadrons both in the vacuum and in the nuclear matter, and have given many
successful descriptions of the hadron  properties \cite{SVZ79}. In Ref.\cite{Wang-NM-B}, we
 study the $\Lambda$-type and $\Sigma$-type heavy baryon states  in the nuclear matter using the QCD sum rules,
 and obtain three coupled QCD sum rules for the masses, vector self-energies and pole residues  in the nuclear matter. In this
 article, we extend out previous works \cite{WangzgEPJA,Wang-NM-B} to study the properties of the  doubly heavy baryon states $\Xi_{cc}$, $\Omega_{cc}$, $\Xi_{bb}$
 and $\Omega_{bb}$  in the nuclear matter.
The mass-shifts of the  heavy mesons, heavy baryons and doubly heavy baryons  differ greatly
 from the corresponding ones of the light mesons and light baryons due to the appearance of the heavy quarks,
 the full propagators of the heavy  quarks in the nuclear matter undergo much slight modifications compared with that of the light quarks,
 the gluon condensates are slightly modified in the nuclear matter.
  The upcoming FAIR (facility for
antiproton and ion research) project at GSI (heavy ion research laboratory)   provides the opportunity to extend the experimental
studies of the hadron properties in the nuclear matter into the charm sector \cite{CBM}, the present  predictions can
 be confronted with the experimental data   in the future.
 The  ground state light-flavor hadrons   in the nuclear matter have been studied extensively with the
QCD sum rules \cite{Drukarev1991,C-parameter,C-parameter-2}, the heavy quarkonia $J/\psi$, $\eta_c$ in the nuclear matter have also
been studied with the QCD sum rules \cite{Jpsi-etac}, while the    works on  the
 heavy mesons   in the nuclear matter   are few \cite{Hayashigaki}.

The article is arranged as follows:  we study the doubly
   heavy baryon states $\Xi_{QQ}$ and $\Omega_{QQ}$ in the nuclear matter
 with  the  QCD sum rules in Sec.2; in Sec.3, we present the
numerical results and discussions; and Sec.4 is reserved for our
conclusions.

\section{The  doubly heavy baryons in the nuclear matter  with  QCD sum rules}
We study the doubly heavy baryon states  $\Xi_{QQ}$ and $\Omega_{QQ}$    in the nuclear matter
with the  two-point correlation functions $\Pi(p)$,
\begin{eqnarray}
\Pi(p) &=& i\int d^4 x\, e^{ip\cdot x}\langle \Psi_0| T\left\{J_q(x)\bar{J}_q(0)\right\}| \Psi_0\rangle \, , \nonumber \\
J_q(x)&=&\epsilon^{ijk} Q^t_i(x)C\gamma_\mu Q_j(x)\gamma^\mu \gamma_5 q_k(x)\, ,
\end{eqnarray}
where the $i$, $j$, $k$ are color indexes, $Q=c,b$, the $C$ is the charge conjunction matrix,
and  the $ |\Psi_0\rangle$ is the nuclear matter ground state. We use the current $J_q(x)$ to interpolate the  $\Xi_{QQ}$,
 and  obtain the current $J_s(x)$ to interpolate the $\Omega_{QQ}$ with a simple replacement $q \to s$.
The correlation functions  $\Pi(p)$ can  be decomposed as
\begin{eqnarray}
\Pi(p)&=& \Pi_s(p^2,p \cdot u)+ \Pi_p(p^2,p \cdot u) \!\not\!{p} +\Pi_u(p^2,p \cdot u)  \!\not\!{u}\,,
\end{eqnarray}
according to Lorentz covariance,  parity and time reversal invariance \cite{Drukarev1991,C-parameter}.
In  the limit four-vector $u_\mu=(1,0)$,  the component $\Pi_i(p^2,p \cdot u)$ reduces  to $\Pi_i(p_0,\vec{p})$,  where $i=s,p,u$.

We  insert a complete set  of intermediate doubly heavy baryon states with the
same quantum numbers as the current operators $J_q(x)$ into the correlation functions $\Pi(p)$ to obtain the hadronic representation
\cite{SVZ79}, then isolate the  ground state contributions of the doubly heavy baryons $\Xi_{QQ}$, and use  the dispersion relation to recast
  the three components  $\Pi_i(p_0,\vec{p})$ into the following form:
\begin{eqnarray}
\Pi_i(p_0,\vec{p})&=&{1\over2\pi i}\int _{-\infty}^\infty~d\omega{\Delta\Pi_i(\omega,\vec{p})\over\omega-p_0} \, ,
\end{eqnarray}
where
\begin{eqnarray}
\Delta\Pi_s(\omega,\vec{p})&=&-2\pi i \frac{\lambda_{\Xi_{QQ}}^{*2}M_{\Xi_{QQ}}^*}{2E^*_p}\left[\delta(\omega-E_p)-\delta(\omega-\overline{E}_p)\right]\,,\nonumber\\
\Delta\Pi_p(\omega,\vec{p})&=&-2\pi i\frac{\lambda_{\Xi_{QQ}}^{*2} }{2E^*_p}\left[\delta(\omega-E_p)-\delta(\omega-\overline{E}_p)\right]\,,\nonumber\\
\Delta\Pi_u(\omega,\vec{p})&=&+2\pi i\frac{\lambda_{\Xi_{QQ}}^{*2}\Sigma^v_{\Xi_{QQ}} }{2E^*_p}\left[\delta(\omega-E_p)-\delta(\omega-\overline{E}_p)\right]\,,
 \end{eqnarray}
 $E_p=\Sigma^v_{\Xi_{QQ}}+E_p^*$, $\overline{E}_p=\Sigma^v_{\Xi_{QQ}}-E_p^*$, $E_p^*= \sqrt{ M_{\Xi_{QQ}}^{*2}+\vec{p}^2}$,
   the $M_{\Xi_{QQ}}^{*}$, $\Sigma^v_{\Xi_{QQ}}$ and $\lambda^*_{\Xi_{QQ}}$ are
 the  masses, vector self-energies and pole residues of the doubly heavy baryon states $\Xi_{QQ}$ respectively in the nuclear matter.

We carry out the operator product expansion in the nuclear matter  at the large space-like region $p^2\ll 0$, and obtain the
  spectral densities at the level of quark-gluon degrees of freedom, then
take the limit $u_\mu=(1,0)$, and obtain the three components $\Pi_i(p_0,\vec{p})$ of the correlation function $\Pi(p)$ \cite{Drukarev1991,C-parameter}:
\begin{eqnarray}
\Pi_i(p_0,\vec{p})&=&\sum_n C_n^i(p_0,\vec{p})\langle{\cal{O}}_n\rangle_{\rho_N}\, ,
\end{eqnarray}
 where the $C_n^i(p_0,\vec{p})$ are the Wilson coefficients,  $\langle{\cal{O}}_n\rangle_{\rho_N}=\langle \Psi_0|{\cal{O}}_n|\Psi_0\rangle=\langle{\cal{O}}_n\rangle+\rho_N\langle
{\cal{O}}_n\rangle_N$  at   low nuclear density in the linear approximation, the   $\langle{\cal{O}}_n\rangle$ and $\langle
{\cal{O}}_n\rangle_N$ denote the vacuum condensates and the nuclear matter induced condensates,  respectively. We can obtain the imaginary parts of the QCD spectral densities
  through the formula
\begin{eqnarray}
\Delta\Pi_i(\omega,\vec{p})&=&{\rm{limit}}_{\epsilon\to 0} \left[\Pi_i(\omega+i\epsilon,\vec{p})-\Pi_i(\omega-i\epsilon,\vec{p}) \right]\, .
\end{eqnarray}

We  match  the hadronic spectral densities with the QCD  spectral densities,
and multiply both sides with  the weight function $(\omega-\overline{E}_p)e^{-\frac{\omega^2}{T^2}}$,
   perform the integral  $\int_{-\omega_0}^{\omega_0}d\omega$,
\begin{eqnarray}
\int_{-\omega_0}^{\omega_0}d\omega\Delta\Pi_i(\omega,\vec{p})(\omega-\overline{E}_p)e^{-\frac{\omega^2}{T^2}}\,,
\end{eqnarray}
to exclude  the negative quasi-particle contributions,
  and obtain the following three QCD sum rules:
  \begin{eqnarray}
\lambda_{\Xi_{QQ}}^{*2}e^{-\frac{E_p^2}{T^2}} &=&\int_{(2m_Q+m_q)^2}^{s_0^*}ds \,\rho^{p}_{\Xi_{QQ}}(s)e^{-\frac{s}{T^2}}\, ,  \\
\lambda_{\Xi_{QQ}}^{*2}M_{\Xi_{QQ}}^*e^{-\frac{E_p^2}{T^2}} &=&\int_{(2m_Q+m_q)^2}^{s_0^*}ds \,\rho^{s}_{\Xi_{QQ}}(s)e^{-\frac{s}{T^2}}\, , \\
\lambda_{\Xi_{QQ}}^{*2}\Sigma^v_{\Xi_{QQ}} e^{-\frac{E_p^2}{T^2}} &=&\int_{(2m_Q+m_q)^2}^{s_0^*}ds \,\rho^{u}_{\Xi_{QQ}}(s)e^{-\frac{s}{T^2}}\, ,
\end{eqnarray}

\begin{eqnarray}
\rho_{\Xi_{QQ}}^{p}(s)  &=&  \frac{3}{8\pi^4}\int_{\alpha_i}^{\alpha_f} d\alpha \int_{\beta_i}^{1-\alpha} d\beta \alpha\beta(1-\alpha-\beta)(s-\widetilde{E}^2_Q)(5s-3\widetilde{E}^2_Q)\nonumber\\
&&+ \frac{3m_Q^2}{8\pi^4}\int_{\alpha_i}^{\alpha_f} d\alpha \int_{\beta_i}^{1-\alpha} d\beta (1-\alpha-\beta)(s-\widetilde{E}^2_Q) \nonumber\\
&&-\frac{m_Q^2}{12\pi^2}\langle\frac{\alpha_sGG}{\pi}\rangle_{\rho_N}\int_{\alpha_i}^{\alpha_f} d\alpha \int_{\beta_i}^{1-\alpha} d\beta
(1-\alpha-\beta)\left[\frac{\alpha}{\beta^2}+\frac{\beta}{\alpha^2} \right]\left[1+\frac{\widetilde{m}_Q^2}{2T^2}\right]\delta(s-\widetilde{E}^2_Q) \nonumber\\
&&-\frac{m_Q^4}{48\pi^2T^2}\langle\frac{\alpha_sGG}{\pi}\rangle_{\rho_N}\int_{\alpha_{i}}^{\alpha_{f}}d\alpha
\int_{\beta_{i}}^{1-\alpha} d\beta (1-\alpha-\beta)\left[\frac{1}{\alpha^3}+\frac{1}{\beta^3}\right]\delta(s-\widetilde{E}^2_Q)\nonumber\\
&&+\frac{m_Q^2}{16\pi^2}\langle\frac{\alpha_sGG}{\pi}\rangle_{\rho_N}\int_{\alpha_{i}}^{\alpha_{f}}d\alpha
\int_{\beta_{i}}^{1-\alpha} d\beta (1-\alpha-\beta)\left[\frac{1}{\alpha^2}+\frac{1}{\beta^2}\right]\delta(s-\widetilde{E}^2_Q)\nonumber\\
&&+\frac{1}{32\pi^2}\langle\frac{\alpha_sGG}{\pi}\rangle_{\rho_N}\int_{\alpha_{i}}^{\alpha_{f}}d\alpha
\int_{\beta_{i}}^{1-\alpha} d\beta \left[3+(1+\alpha+\beta)\widetilde{m}_Q^2\delta(s-\widetilde{E}^2_Q)\right]\nonumber\\
&&+\frac{m_q\langle \bar{q} q\rangle_{\rho_N}}{3\pi^2}\int_{\alpha_{i}}^{\alpha_{f}}d\alpha \alpha(1-\alpha)\left[5+(3\widehat{m}_Q^2-2\widehat{E}_Q^2)\delta(s-\widehat{E}_Q^2)\right]\nonumber\\
&&-\frac{\langle q^{\dagger}iD_0 q\rangle_{\rho_N}}{3\pi^2}\int_{\alpha_{i}}^{\alpha_{f}}d\alpha \alpha(1-\alpha)\left[2+(3\widehat{m}_Q^2-8\widehat{E}_Q^2)\delta(s-\widehat{E}_Q^2)\right]\nonumber\\
&&-\overline{E}_p \left\{ -\frac{\langle q^\dagger q\rangle_{\rho_N}}{\pi^2}\int_{\alpha_{i}}^{\alpha_{f}}d\alpha \alpha(1-\alpha)
+\frac{\langle q^{\dagger}g_s\sigma G q\rangle_{\rho_N}}{3\pi^2}\int_{\alpha_{i}}^{\alpha_{f}}d\alpha \alpha(1-\alpha)\left[\frac{5}{2}+\frac{\widehat{m}_Q^2-\widehat{E}_Q^2}{T^2}\right] \right.\nonumber\\
&& \delta(s-\widehat{E}_Q^2)+\frac{\langle q^{\dagger}iD_0iD_0 q\rangle_{\rho_N}}{\pi^2}\int_{\alpha_{i}}^{\alpha_{f}}d\alpha \alpha(1-\alpha)\left[4+\frac{8\widehat{m}_Q^2-12\widehat{E}_Q^2}{3T^2}\right]\delta(s-\widehat{E}_Q^2) \nonumber\\
&& \left.-\frac{\langle q^{\dagger}g_s\sigma G q\rangle_{\rho_N}}{12\pi^2}\int_{\alpha_{i}}^{\alpha_{f}}d\alpha \delta(s-\widehat{E}_Q^2)\right\}\, ,
\end{eqnarray}

\begin{eqnarray}
\rho_{\Xi_{QQ}}^{s}(s)  &=&  \frac{3m_q}{8\pi^4}\int_{\alpha_i}^{\alpha_f} d\alpha \int_{\beta_i}^{1-\alpha} d\beta \alpha\beta(s-\widetilde{E}^2_Q)(2s-\widetilde{E}^2_Q)+ \frac{3m_q m_Q^2}{4\pi^4}\int_{\alpha_i}^{\alpha_f} d\alpha \int_{\beta_i}^{1-\alpha} d\beta (s-\widetilde{E}^2_Q) \nonumber\\
&&-\frac{m_q m_Q^2}{48\pi^2}\langle\frac{\alpha_sGG}{\pi}\rangle_{\rho_N}\int_{\alpha_i}^{\alpha_f} d\alpha \int_{\beta_i}^{1-\alpha} d\beta
 \left[\frac{\alpha}{\beta^2}+\frac{\beta}{\alpha^2} \right]\left[1+\frac{\widetilde{m}_Q^2}{T^2}\right]\delta(s-\widetilde{E}^2_Q) \nonumber\\
&&-\frac{m_q m_Q^4}{24\pi^2T^2}\langle\frac{\alpha_sGG}{\pi}\rangle_{\rho_N}\int_{\alpha_{i}}^{\alpha_{f}}d\alpha
\int_{\beta_{i}}^{1-\alpha} d\beta  \left[\frac{1}{\alpha^3}+\frac{1}{\beta^3}\right]\delta(s-\widetilde{E}^2_Q)\nonumber\\
&&+\frac{m_q m_Q^2}{8\pi^2}\langle\frac{\alpha_sGG}{\pi}\rangle_{\rho_N}\int_{\alpha_{i}}^{\alpha_{f}}d\alpha
\int_{\beta_{i}}^{1-\alpha} d\beta  \left[\frac{1}{\alpha^2}+\frac{1}{\beta^2}\right]\delta(s-\widetilde{E}^2_Q)\nonumber\\
&&-\frac{m_q }{16\pi^2}\langle\frac{\alpha_sGG}{\pi}\rangle_{\rho_N}\int_{\alpha_{i}}^{\alpha_{f}}d\alpha
\int_{\beta_{i}}^{1-\alpha} d\beta  \left[ 1+\frac{\widetilde{m}_Q^2}{2}\delta(s-\widetilde{E}^2_Q) \right]\nonumber\\
&&-\frac{\langle \bar{q} q\rangle_{\rho_N}}{2\pi^2}\int_{\alpha_{i}}^{\alpha_{f}}d\alpha \alpha(1-\alpha)\left[3s+2\widehat{m}_Q^2-2\widehat{E}_Q^2   \right]\nonumber\\
&&+\frac{\langle \bar{q}g_s\sigma G q\rangle_{\rho_N}}{ 2\pi^2}\int_{\alpha_{i}}^{\alpha_{f}}d\alpha \alpha(1-\alpha)\left[\frac{3}{2}+\left(\frac{5\widehat{m}_Q^2}{2}- \widehat{E}_Q^2 +\frac{\widehat{m}_Q^4}{T^2}-\frac{\widehat{m}_Q^2\widehat{E}_Q^2}{T^2}\right)\delta(s-\widehat{E}_Q^2)\right]\nonumber\\
&&  +\frac{\langle \bar{q}iD_0iD_0 q\rangle_{\rho_N}}{\pi^2}\int_{\alpha_{i}}^{\alpha_{f}}d\alpha \alpha(1-\alpha)\left[4\widehat{m}_Q^2-4\widehat{E}_Q^2+\frac{\widehat{m}_Q^4}{T^2}-\frac{4\widehat{m}_Q^2\widehat{E}_Q^2}{T^2}\right]\delta(s-\widehat{E}_Q^2) \nonumber\\
&&-\frac{\langle \bar{q}g_s\sigma G q\rangle_{\rho_N}}{ 2\pi^2}\int_{\alpha_{i}}^{\alpha_{f}}d\alpha  \left[\frac{1}{4}+\left(\frac{ \widehat{m}_Q^2}{3}-\frac{\widehat{E}_Q^2}{6}\right)\delta(s-\widehat{E}_Q^2)\right]\nonumber\\
&&+\frac{m_q\langle \bar{q}iD_0 q\rangle_{\rho_N}-\langle \bar{q}iD_0iD_0 q\rangle_{\rho_N}}{ 6\pi^2}\int_{\alpha_{i}}^{\alpha_{f}}d\alpha
\left[  4\widehat{E}_Q^2 -\widehat{m}_Q^2\right]\delta(s-\widehat{E}_Q^2)\nonumber\\
&&-\overline{E}_p \left\{ \frac{3m_q\langle q^\dagger q\rangle_{\rho_N}}{\pi^2}\int_{\alpha_{i}}^{\alpha_{f}}d\alpha \alpha(1-\alpha)\left[1+\widehat{m}_Q^2\delta(s-\widehat{E}_Q^2)\right]  \right\} \, ,
\end{eqnarray}

\begin{eqnarray}
\rho_{\Xi_{QQ}}^{u}(s)  &=& \frac{\langle  q^{\dagger} q\rangle_{\rho_N}}{2\pi^2}\int_{\alpha_{i}}^{\alpha_{f}}d\alpha \alpha(1-\alpha)\left[s+\widehat{m}_Q^2-\widehat{E}_Q^2   \right]\nonumber\\
&&-\frac{\langle q^{\dagger}g_s\sigma G q\rangle_{\rho_N}}{  2\pi^2}\int_{\alpha_{i}}^{\alpha_{f}}d\alpha \alpha(1-\alpha)\left[\frac{1}{2}+\left(\frac{5\widehat{m}_Q^2}{6}-\widehat{E}_Q^2+\frac{\widehat{m}_Q^4}{3T^2}-\frac{\widehat{m}_Q^2\widehat{E}_Q^2}{3T^2}\right)\delta(s-\widehat{E}_Q^2)\right]\nonumber\\
&&  -\frac{\langle \bar{q}iD_0iD_0 q\rangle_{\rho_N}}{\pi^2}\int_{\alpha_{i}}^{\alpha_{f}}d\alpha \alpha(1-\alpha)\left[\frac{8\widehat{m}_Q^2}{3}-6\widehat{E}_Q^2+\frac{\widehat{m}_Q^4}{3T^2}-\frac{2\widehat{m}_Q^2\widehat{E}_Q^2}{T^2}\right]\delta(s-\widehat{E}_Q^2) \nonumber\\
&&+\frac{\langle q^{\dagger}g_s\sigma G q\rangle_{\rho_N}}{  2\pi^2}\int_{\alpha_{i}}^{\alpha_{f}}d\alpha \left[\frac{1}{4}+ \frac{\widehat{m}_Q^2}{3}\delta(s-\widehat{E}_Q^2)\right]\nonumber\\
&&-\overline{E}_p \left\{  \frac{ m_q\langle \bar{q} q\rangle_{\rho_N}-4\langle q^\dagger iD_0 q\rangle_{\rho_N}}{3\pi^2}\int_{\alpha_{i}}^{\alpha_{f}}d\alpha \alpha(1-\alpha)\left[2+\widehat{m}_Q^2\delta(s-\widehat{E}_Q^2)\right]     \right\} \, ,
\end{eqnarray}
where $\widetilde{E}_Q^2=\frac{m_Q^2}{\alpha}+\frac{m_Q^2}{\beta}+\vec{p}^2$, $\widetilde{m}_Q^2=\frac{m_Q^2}{\alpha}+\frac{m_Q^2}{\beta}$,
$ \widehat{E}_Q^2=\frac{m_Q^2}{\alpha(1-\alpha)} +\vec{p}^2$, $ \widehat{m}_Q^2=\frac{m_Q^2}{\alpha(1-\alpha)} $, $\alpha_{f}=\frac{1+\sqrt{1-4m_Q^2/s}}{2}$,
$\alpha_{i}=\frac{1-\sqrt{1-4m_Q^2/s}}{2}$,
$\beta_{i}=\frac{\alpha m_Q^2}{\alpha s -m_Q^2}$, $s^*_0=\omega_0^2=s^0_{\Xi_{QQ}}-\vec{p}^2$,
  the $s^0_{\Xi_{QQ}}$ are the continuum threshold parameters, the $T^2$ are the Borel
parameters,  $\alpha_i, \beta_i\rightarrow 0$, $\alpha_f \rightarrow 1$  in the spectral densities where the $\delta$ functions $\delta(s-\widetilde{E}_Q^2)$ and  $\delta(s-\widehat{E}_Q^2)$ appear.
We can obtain the  masses $M^*_{\Xi_{QQ}}$, vector
self-energies $\Sigma^v_{\Xi_{QQ}}$ and pole residues $\lambda^*_{\Xi_{QQ}}$ in the nuclear matter by solving the three
equations Eqs.(8-10) with simultaneous  iterations.
With the simple replacements of the corresponding  parameters, such as  $m_q \to m_s$, $\langle\bar{q}q\rangle_{\rho_N} \to \langle\bar{s}s\rangle_{\rho_N}$,
$\langle q^{\dagger}q\rangle_{\rho_N} \to \langle s^{\dagger}s\rangle_{\rho_N}$, etc,
   we can obtain the corresponding three QCD sum rules for the doubly heavy baryon states $\Omega_{QQ}$.

\section{Numerical results and discussions}
 The input parameters are taken  as
 $\langle q^\dagger q\rangle_{\rho_N}={3\over2}\rho_N$, $\langle s^\dagger s\rangle_{\rho_N}=0$,
 $\langle \bar{q} q\rangle_{\rho_N}=\langle \bar{q} q\rangle+{\sigma_N\over m_u+m_d}\rho_N $,
  $\langle \bar{s} s\rangle_{\rho_N}=\langle \bar{s} s\rangle+y{\sigma_N\over m_u+m_d}\rho_N $,
  $m_u+m_d=12\,\rm{MeV}$,  $\sigma_N=(45\pm 10)\,\rm{MeV}$,
 $\langle\frac{\alpha_sGG}{\pi}\rangle_{\rho_N}=\langle\frac{\alpha_sGG}{\pi}\rangle-(0.65\pm0.15)\,{\rm GeV}\rho_N$,
 $\langle\frac{\alpha_sGG}{\pi}\rangle=(0.33\,\rm{GeV})^4$,
 $\langle q^\dagger iD_0 q\rangle_{\rho_N}=0.18\,{\rm GeV}\rho_N$,
 $\langle s^\dagger iD_0 s\rangle_{\rho_N}=\frac{m_s}{4}\langle \bar{s} s\rangle_{\rho_N}+0.02\,{\rm GeV}\rho_N$,
 $\langle q^{\dagger} iD_0iD_0 q\rangle_{\rho_N}+{1 \over 12}\langle q^{\dagger}g_s\sigma G q\rangle_{\rho_N}=0.031\,{\rm{GeV}}^2\rho_N$,
$\langle s^{\dagger} iD_0iD_0 s\rangle_{\rho_N}+{1 \over 12}\langle s^{\dagger}g_s\sigma G s\rangle_{\rho_N}=y0.031\,{\rm{GeV}}^2\rho_N$,
$\langle\bar{q}g_s\sigma Gq\rangle=m_0^2\langle\bar{q}q\rangle$, $\langle\bar{s}g_s\sigma Gs\rangle=m_0^2\langle\bar{s}s\rangle$,
$\langle \bar{q} iD_0iD_0 q\rangle_{\rho_N}+{1\over 8}\langle\bar{q}g_s\sigma G q\rangle_{\rho_N}=0.3\,{\rm{GeV}}^2\rho_N$,
$\langle \bar{s} iD_0iD_0 s\rangle_{\rho_N}+{1\over 8}\langle\bar{s}g_s\sigma G s\rangle_{\rho_N}=y0.3\,{\rm{GeV}}^2\rho_N$,
$\langle\bar{q}g_s\sigma G q\rangle_{\rho_N}=\langle\bar{q}g_s\sigma G q\rangle+3.0\,{\rm GeV}^2\rho_N$,
$\langle\bar{s}g_s\sigma G s\rangle_{\rho_N}=\langle\bar{s}g_s\sigma G s\rangle+y3.0\,{\rm GeV}^2\rho_N$,
$\langle q^{\dagger}g_s\sigma G q\rangle_{\rho_N}=-0.33\,{\rm GeV}^2\rho_N$, $\langle s^{\dagger}g_s\sigma G s\rangle_{\rho_N}=-y0.33\,{\rm GeV}^2\rho_N$,
  $\langle\bar{q}q\rangle=-(0.23\,\rm{GeV})^3$, $\langle\bar{s}s\rangle=0.8\langle\bar{q}q\rangle$, $m_0^2=0.8\,\rm{GeV}^2$,
$\rho_N=(0.11\,\rm{GeV})^3$,   $y=0.3\pm0.3$, $m_q=6\,\rm{MeV}$,  $m_s=0.13\,\rm{GeV}$, $m_c=1.35\,\rm{GeV}$ and
$m_b=4.7\,\rm{GeV}$ at the
energy scale  $\mu=1\, \rm{GeV}$ \cite{C-parameter,C-parameter-2}.

We can recover the QCD sum rules in the vacuum by taking the limit  $\rho_N=0$, then differentiate  the Eqs.(8-9) with respect to  $\frac{1}{T^2}$ respectively,
and eliminate the pole residues $\lambda_{\Xi_{QQ}}$ (here we smear the  asterisk $*$ to denote the pole residues in the vacuum),
and obtain  two QCD sum rules for  the masses  $M_{\Xi_{QQ}}$ with respect to the spinor structures $ \!\not\!{p}$ and 1, respectively. On the other hand, we can divide Eq.(9) by Eq.(8) to obtain the QCD sum rules
for the masses $M_{\Xi_{QQ}}$.  From Eq.(10), we can see that if we take the limit
$\rho_N=0$,   $\Sigma^v_{\Xi_{QQ}}\neq0$, due to  appearance of the term $m_q\langle \bar{q} q\rangle_{\rho_N}$, however, such terms are small enough to be
  neglected safely. The QCD spectral densities from the operator product expansion in the vacuum and in nuclear matter by taking the limit $\rho_N=0$ have slight
discrepancies, the discrepancies cannot result in remarkable differences  on the masses and pole residues.

In Ref.\cite{WangzgEPJA},  we study the doubly heavy baryon
states  $\Xi_{QQ}$ and $\Omega_{QQ}$    by subtracting the possible
contributions from the corresponding negative-parity doubly heavy
baryon states with the QCD sum rules. In that reference, we choose the spinor structure $1+\gamma^0$,
take the Borel parameters  as $T^2_{\Xi_{cc}}=(2.8-3.8)\,\rm{GeV}^2$,
$T^2_{\Omega_{cc}}=(3.0-4.0)\,\rm{GeV}^2$, $T^2_{\Xi_{bb}}=(7.7-9.1)\,\rm{GeV}^2$, $T^2_{\Omega_{bb}}=(7.9-9.3)\,\rm{GeV}^2$, the
continuum threshold parameters as
$ s^0_{\Xi_{cc}}=(4.2\pm0.1\,\rm{GeV})^2$, $ s^0_{\Omega_{cc}}=(4.3\pm0.1\,\rm{GeV})^2$, $ s^0_{\Xi_{bb}}=(10.8\pm0.1\,\rm{GeV})^2$,
$s^0_{\Omega_{bb}}=(10.9\pm0.1\,\rm{GeV})^2$, and obtain the masses $M_{\Xi_{cc}}=(3.57 \pm 0.14)\,\rm{GeV}$,
$M_{\Omega_{cc}}=(3.71 \pm 0.14)\,\rm{GeV}$, $M_{\Xi_{bb}}=(10.17 \pm 0.14)\,\rm{GeV}$, $M_{\Omega_{bb}}=(10.32 \pm 0.14)\,\rm{GeV}$.
The predictions  are consistent with the value from the
   SELEX collaboration $M_{\Xi_{cc}}=(3518.9 \pm 0.9) \,\rm{ MeV }$ \cite{SELEX2002,SELEX2004}. For the technical details, one can consult Ref.\cite{WangzgEPJA}.

   In this article, we choose the spin structures $1$, $\!\not\!{p}$ and $\!\not\!{u}$, see Eq.(2),  take the continuum threshold parameters as $s^0_{\Xi_{cc}}=(17.5\pm0.5)\,\rm{GeV}^2$,
 $s^0_{\Omega_{cc}}=(18.0\pm0.5)\,\rm{GeV}^2$,  $s^0_{\Xi_{bb}}=(117.0\pm1.0)\,\rm{GeV}^2$ and
 $s^0_{\Omega_{bb}}=(119.0\pm1.0)\,\rm{GeV}^2$  consulting  Ref.\cite{WangzgEPJA}, and vary the Borel parameters $T^2$ to reproduce almost
 the same masses as Ref.\cite{WangzgEPJA}. In calculations, the Borel parameters are taken as $T_{\Xi_{cc}}^2=(2.4-2.8)\,\rm{GeV}^2$, $T_{\Omega_{cc}}^2=(2.4-2.8)\,\rm{GeV}^2$, $T_{\Xi_{bb}}^2=(7.0-7.5)\,\rm{GeV}^2$ and $T_{\Omega_{bb}}^2=(6.7-7.2)\,\rm{GeV}^2$, respectively.

The interpolating  currents $J_q(x)$ have nonvanishing couplings to both the positive- and negative-parity doubly heavy baryon states. For the QCD sum rules in the nuclear matter,
we cannot choose the spinor structure $1+\gamma^0$ to avoid the possible contaminations from the negative-parity  doubly heavy baryon states. In this article, we
choose the spinor structures $1$ and $\!\not\!{p}$ in stead of the spinor structure $1+\gamma^0$  in
the limit $\rho_N=0$, the ratios between the negative- and  positive-parity contributions  are about $-(0-9)\%$  and $-(4-8)\%$ in the charm and bottom sectors,
 respectively \cite{WangzgEPJA}. For the technical details, one can consult Ref.\cite{WangzgEPJA}. The
contaminations from the negative-parity doubly heavy baryon states
are rather small. We can also take into account the contributions from the negative-parity doubly heavy baryon states explicitly as Ref.\cite{K-SR}, and perform detailed analysis,
 the tedious analysis maybe our
next work.

   If we take the limit $m_u= m_d=m_q\rightarrow0$ in the QCD sum rules for the nucleons,
   the  hadronic parameters $ M_N^*$ and $\Sigma^v_{N}$ in the nuclear matter   can be approximated as
   \begin{eqnarray}
 M_N^*&=&-\frac{8\pi^2}{T^2}\langle\bar{q}q\rangle_{\rho_N} \, , \nonumber\\
  \Sigma^v_N&=&\frac{64\pi^2}{3T^2}\langle q^{\dagger}q\rangle_{\rho_N}\, ,
  \end{eqnarray}
respectively. Such relations are simple and elegant, however, the predictions change monotonously with  variations of the Borel parameter $T^2$.
In the present case, if we take the limit $m_q\rightarrow0$ and neglect other terms of minor importance, we can obtain the following relations,
\begin{eqnarray}
 M_{\Xi_{QQ}}^*&=&f(T^2,m_Q,s^0_{\Xi_{QQ}})\langle\bar{q}q\rangle_{\rho_N} \, ,\nonumber\\
  \Sigma^v_{\Xi_{QQ}}&=&g(T^2,m_Q,s^0_{\Xi_{QQ}}) \langle q^{\dagger}q\rangle_{\rho_N} \, ,
  \end{eqnarray}
from Eqs.(8-13). The formal notations $f(T^2,m_Q,s^0_{\Xi_{QQ}})$ and $g(T^2,m_Q,s^0_{\Xi_{QQ}})$ are complex functions of
the variables $T^2$, $m_Q$ and $s^0_{\Xi_{QQ}}$. It is difficult to carry out the integrals  over the variables
$s$, $\alpha$, $\beta$   analytically due to the appearance of the heavy quark mass $m_Q$, we expect
that the $f(T^2,m_Q,s^0_{\Xi_{QQ}})$ and $g(T^2,m_Q,s^0_{\Xi_{QQ}})$ are sensitive  to the Borel parameter $T^2$,
the numerical calculations confirm  such conjecture.

 Finally, we obtain the masses and pole residues in the vacuum
  $M_{\Xi_{cc}}=(3.53\pm0.32)\,\rm{GeV}$, $M_{\Omega_{cc}}=(3.61\pm0.29)\,\rm{GeV}$,  $M_{\Xi_{bb}}=(10.14\pm0.47)\,\rm{GeV}$,
  $M_{\Omega_{bb}}=(10.27\pm0.46)\,\rm{GeV}$, $\lambda_{\Xi_{cc}}=(0.117\pm0.046)\,\rm{GeV}$, $\lambda_{\Omega_{cc}}=(0.128\pm0.047)\,\rm{GeV}$,
$\lambda_{\Xi_{bb}}=(0.287\pm0.162)\,\rm{GeV}$, $\lambda_{\Omega_{bb}}=(0.359\pm0.211)\,\rm{GeV}$; the masses, pole residues and vector self-energies in the nuclear matter   $M_{\Xi_{cc}}^{*}=(2.43\pm0.27)\,\rm{GeV}$, $M_{\Omega_{cc}}^{*}=(3.28\pm0.26)\,\rm{GeV}$,  $M_{\Xi_{bb}}^{*}=(6.77\pm0.38)\,\rm{GeV}$,
  $M_{\Omega_{bb}}^{*}=(9.22\pm0.41)\,\rm{GeV}$, $\lambda_{\Xi_{cc}}^{*}=(0.033\pm0.005)\,\rm{GeV}$,
  $\lambda_{\Omega_{cc}}^{*}=(0.078\pm0.021)\,\rm{GeV}$,  $\lambda_{\Xi_{bb}}^{*}=(0.008\pm0.002)\,\rm{GeV}$,
   $\lambda_{\Omega_{bb}}^{*}=(0.075\pm0.033)\,\rm{GeV}$,   $\Sigma_{\Xi_{cc}}^{v}=(0.139\pm0.005)\,\rm{GeV}$,
   $\Sigma_{\Omega_{cc}}^{v}=-(0.006\pm0.001 )\,\rm{GeV}$, $\Sigma_{\Xi_{bb}}^{v}=(0.507\pm0.019)\,\rm{GeV}$, $\Sigma_{\Omega_{bb}}^{v}=(0.009\pm0.001 )\,\rm{GeV}$.
     The mass-shifts in the nuclear matter are $\Delta M_{\Xi_{cc}}=-(1.11\pm0.04)\,\rm{GeV}$,
$\Delta M_{\Omega_{cc}}=-(0.33\pm0.03)\,\rm{GeV}$,
$\Delta M_{\Xi_{bb}}=-(3.37\pm0.09)\,\rm{GeV}$,
$\Delta M_{\Omega_{bb}}=-(1.05\pm0.06)\,\rm{GeV}$, where the mass-shifts $\Delta M$  defined by $\Delta M=M^*-M$  are the scalar self-energies.
The uncertainties come from the Borel parameters $T^2$.

The mass modifications in the nuclear matter are about $-31\%$,  $-33\%$, $-9\%$ and $-10\%$ for the doubly heavy baryon states $\Xi_{cc}$,
$\Xi_{bb}$, $\Omega_{cc}$ and $\Omega_{bb}$, respectively, which are consistent with the quark condensates modifications in the nuclear matter
$\frac{\langle\bar{q}q\rangle_{\rho_N}-\langle\bar{q}q\rangle}{\langle\bar{q}q\rangle}\approx-41\%$,
 $\frac{\langle\bar{s}s\rangle_{\rho_N}-\langle\bar{s}s\rangle}{\langle\bar{s}s\rangle}\approx-(15\pm15)\%$, where the uncertainty
 $\pm 15\%$ comes from the uncertainty of the  parameter $\delta y= \pm0.3$. There appear additional quark condensates associated with the light flavor quarks, such as
 $\langle q^{\dagger}q\rangle_{\rho_N}$, $\langle q^\dagger iD_0 q\rangle_{\rho_N}$, $\langle q^\dagger iD_0 iD_0q\rangle_{\rho_N}$,
 $\langle \bar{q}  iD_0 iD_0q\rangle_{\rho_N}$, etc, their contributions are not large but  cannot be  neglected safely. The $\Xi_{QQ}$ and $\Omega_{QQ}$
 baryons have two heavy quarks besides a light quark,
  the heavy quark  interacts  with the nuclear matter through  the exchange of the intermediate gluons,  and
  the modifications of the gluon condensates in the nuclear matter are mild, $\langle \frac{\alpha_sGG}{\pi}\rangle_{\rho_N}\approx0.93\langle \frac{\alpha_sGG}{\pi}\rangle$ ,  we  expect that
 the mass modifications  are smaller than that of the nucleons, which have three light quarks, $uud$ or $udd$, and the approximation in Eq.(15) makes sense.

  If we take into account the uncertainties of the continuum threshold parameters,
  $\delta s^0_{\Xi_{cc}}=\pm0.5\,\rm{GeV}^2$,  $\delta s^0_{\Omega_{cc}}=\pm0.5\,\rm{GeV}^2$,
   $\delta s^0_{\Xi_{bb}}=\pm1.0\,\rm{GeV}^2$, $\delta s^0_{\Omega_{bb}}=\pm1.0\,\rm{GeV}^2$,
   additional uncertainties
  $\delta \Delta M_{\Xi_{cc}}=\pm 0.03\,\rm{GeV}$,
    $\delta \Delta M_{\Omega_{cc}}=\pm 0.01\,\rm{GeV}$,
   $\delta \Delta M_{\Xi_{bb}}=\pm 0.10\,\rm{GeV}$,
    $\delta \Delta M_{\Omega_{bb}}=\pm 0.04\,\rm{GeV}$ for the mass-shifts $\Delta M$ are introduced. The uncertainty $\delta y=\pm0.3$ leads to the
    uncertainties  $\delta\Delta  M_{\Omega_{cc}}=\pm 0.43\,\rm{GeV}$ and $\delta \Delta M_{\Omega_{bb}}=\pm 1.13\,\rm{GeV}$. The uncertainty $\delta \sigma_N=10\,\rm{MeV}$
    leads to the uncertainties $\delta \Delta M_{\Xi_{cc}}=\pm 0.37\,\rm{GeV}$,
    $\delta \Delta M_{\Omega_{cc}}=\pm 0.10\,\rm{GeV}$,
   $\delta  \Delta M_{\Xi_{bb}}=\pm 1.05\,\rm{GeV}$ and
    $\delta \Delta M_{\Omega_{bb}}=\pm 0.27\,\rm{GeV}$, respectively.

  We can also take into account the perturbative $\mathcal {O}(\alpha_s^n)$ corrections   in the leading logarithmic
approximation through multiplying the QCD spectral densities by the anomalous-dimension factors,
    \begin{eqnarray}
    \left[\frac{\log(T^2/\Lambda^2_{QCD})}{\log(\mu^2/\Lambda^2_{QCD})}\right]^{-2\Gamma_J+\Gamma_{\mathcal{O}}} \, ,
     \end{eqnarray}
       where the anomalous-dimensions of the interpolating  currents are $\Gamma_{J}=\frac{6}{33-2N_f}$, and the anomalous-dimensions of the local operators $\mathcal{O}$ are
       $\Gamma_{\langle\bar{q}q\rangle}=\frac{12}{33-2N_f}$,
     $\Gamma_{m_q\langle\bar{q}q\rangle}=0$, $\Gamma_{\langle g_s^2GG\rangle}=0$, and the $N_f$ is the flavor number.
 The anomalous-dimension factors can lead to the uncertainties   $\frac{\delta M_{\Xi_{cc}}}{M_{\Xi_{cc}}}=(13-17)\%$,
 $\frac{\delta M_{\Omega_{cc}}}{M_{\Omega_{cc}}}=(10-13)\%$, $\frac{\delta M_{\Xi_{bb}}}{M_{\Xi_{bb}}}=(28-36)\%$,
 $\frac{\delta M_{\Omega_{bb}}}{M_{\Omega_{bb}}}=(21-27)\%$,
    $\frac{\delta M_{\Xi_{cc}}^*}{M_{\Xi_{cc}}^*}=(16-21)\%$,
    $\frac{\delta M_{\Omega_{cc}}^*}{M_{\Omega_{cc}}^*}=(10-13)\%$,
      $\frac{\delta M_{\Xi_{bb}}^*}{M_{\Xi_{bb}}^*}=(34-46)\%$,
         $\frac{\delta M_{\Omega_{bb}}^*}{M_{\Omega_{bb}}^*}=(20-26)\%$  for the values
       $\Lambda_{QCD}=(200-300)\,\rm{MeV}$. The masses in the vacuum and in the nuclear matter are increased significantly,  we can
       normalize the masses in the vacuum to the values $M_{\Xi_{cc}}=3.53\,\rm{GeV}$, $M_{\Omega_{cc}}=3.61\,\rm{GeV}$,  $M_{\Xi_{bb}}=10.14\,\rm{GeV}$,
  $M_{\Omega_{bb}}=10.27\,\rm{GeV}$, and estimate  the uncertainties of the mass-shifts due to the energy scales as
  $\frac{\delta \Delta M_{\Xi_{cc}}}{\Delta M_{\Xi_{cc}}}=-(5-7)\%$, $\frac{\delta \Delta M_{\Omega_{cc}}}{\Delta M_{\Omega_{cc}}}=(4-5)\%$,
$\frac{\delta \Delta M_{\Xi_{bb}}}{\Delta M_{\Xi_{bb}}}=-(10-14)\%$, $\frac{\delta \Delta M_{\Omega_{bb}}}{\Delta M_{\Omega_{bb}}}=(6-8)\%$.

     We can refit the Borel parameters and threshold parameters to reproduce the experimental   data approximately \cite{SELEX2004}, the Borel parameters
     (or the effectively energy scales) and the
     threshold parameters have some correlations. In calculations, we observe that larger threshold parameters can lead to smaller masses and cancel out the enhanced
      factors induced  by the non-zero anomalous-dimensions in Eq.(16), however, the contributions from the high resonances and continuum states are included in. So we can
      fix the threshold parameters and take larger Borel parameters to cancel out those  enhanced
      factors, see Eqs.(14-15), then the uncertainties of the mass-shifts induced by the energy scales are about a few percents, and can be neglected safely.
      On the other hand, we can   introduce Borel parameter dependent threshold
     parameters and  study the systematic uncertainties \cite{ChSh3}, it is a hard work before the experimental data are enough and the precise values of the pole residues are known,  we postpone  those works in the future.

If we take the Ioffe current to interpolate the proton,  the QCD sum rules indicate that there exists  a positive
vector self-energy  $\Sigma_N^v=(0.23-0.35)\,\rm{GeV}$ with the typical values of the relevant condensates
and other input parameters and a reasonable negative scalar self-energy  with the suitable parameters
 \cite{C-parameter}.
There exists substantial cancelation between the scalar and vector self-energies,
the  self-energies $\Sigma_N^s$ and $\Sigma_N^v$, which  correspond to the real energy-independent optical potentials $S$ and $V$, satisfy the relation
  ${\Sigma_N^s}/{\Sigma_N^v}\approx -1$ in the leading order approximation.
While the mean-field models  predicate  that  the typical self-energies of the nucleons  in nuclear matter saturation density are
$\Sigma_N^s\approx -350 \,\rm{ MeV}$ and $\Sigma_N^v\approx +300 \,\rm{MeV}$ respectively,   the effective non-relativistic central
potentials $S+V$ are about tens of $\rm{MeV}$.  In the present case,  $\Sigma^s_{\Xi_{cc}}+\Sigma^v_{\Xi_{cc}}=-0.97\,\rm{GeV}$, $\Sigma^s_{\Omega_{cc}}+\Sigma^v_{\Omega_{cc}}=-0.34\,\rm{GeV}$,
  $\Sigma^s_{\Xi_{bb}}+\Sigma^v_{\Xi_{bb}}=-2.86\,\rm{GeV}$, $\Sigma^s_{\Omega_{bb}}+\Sigma^v_{\Omega_{bb}}=-1.04\,\rm{GeV}$, the net optical potentials $S+V$ are large,
  as the self-energies $|\Sigma^v|\ll |\Sigma^s|$.
  The present prediction of the mass-shift $\delta M_{\Xi_{cc}}=-1.11\,\rm{GeV}$ can be confronted with the experimental data  from the
   CBM and $\rm{\bar{P}ANDA}$ collaborations in the future \cite{CBM}, where
 the  properties of the charmed baryons in the nuclear matter will be studied. The $\Xi_{cc}$ have interesting properties, for example,
the theoretical predictions of their lifetimes based on different quark models are in agreement  with each other \cite{Chang0704},
but about one-order larger than the upper limit of the experimental data \cite{SELEX2002}. The precise measurement of the lifetime by
the $\rm{\bar{P}ANDA}$ and LHCb collaborations in the future
 maybe shed light on the apparent discrepancy.

\begin{figure}
 \centering
 \includegraphics[totalheight=5cm,width=7cm]{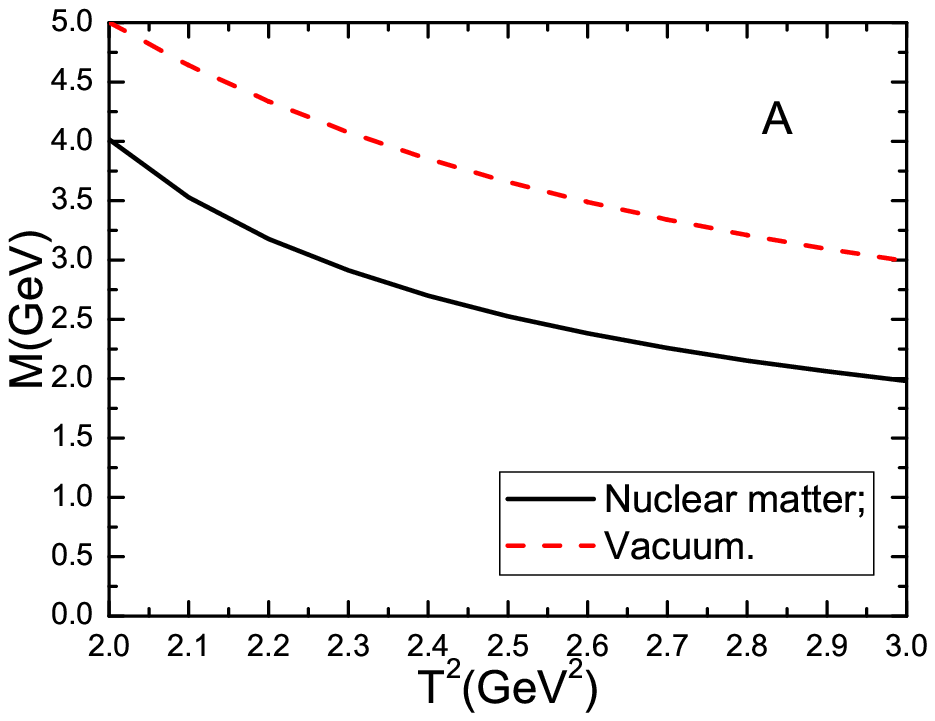}
 \includegraphics[totalheight=5cm,width=7cm]{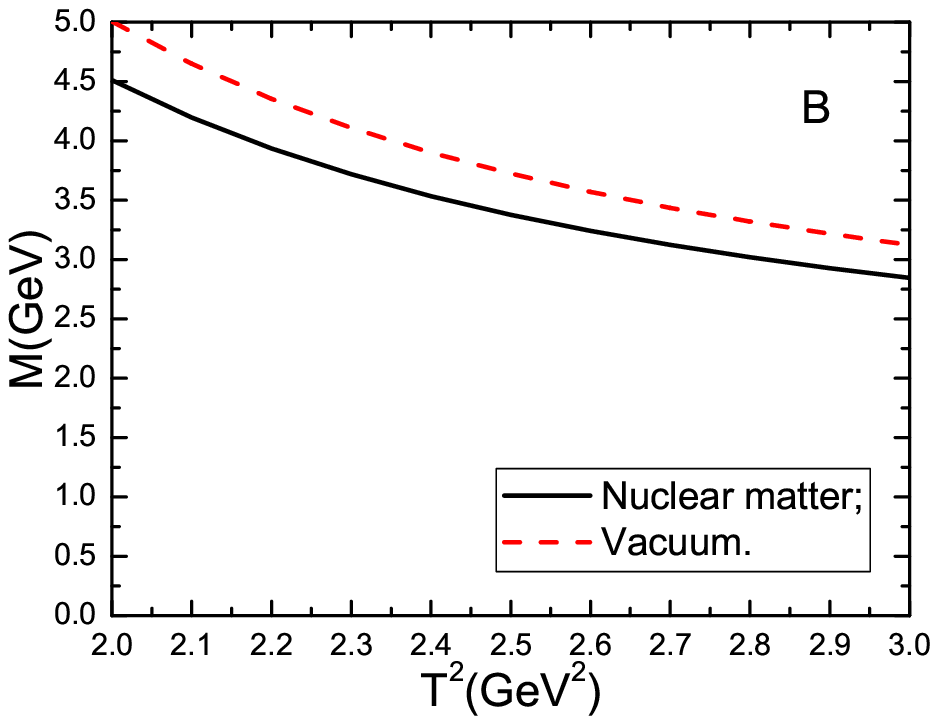}
 \includegraphics[totalheight=5cm,width=7cm]{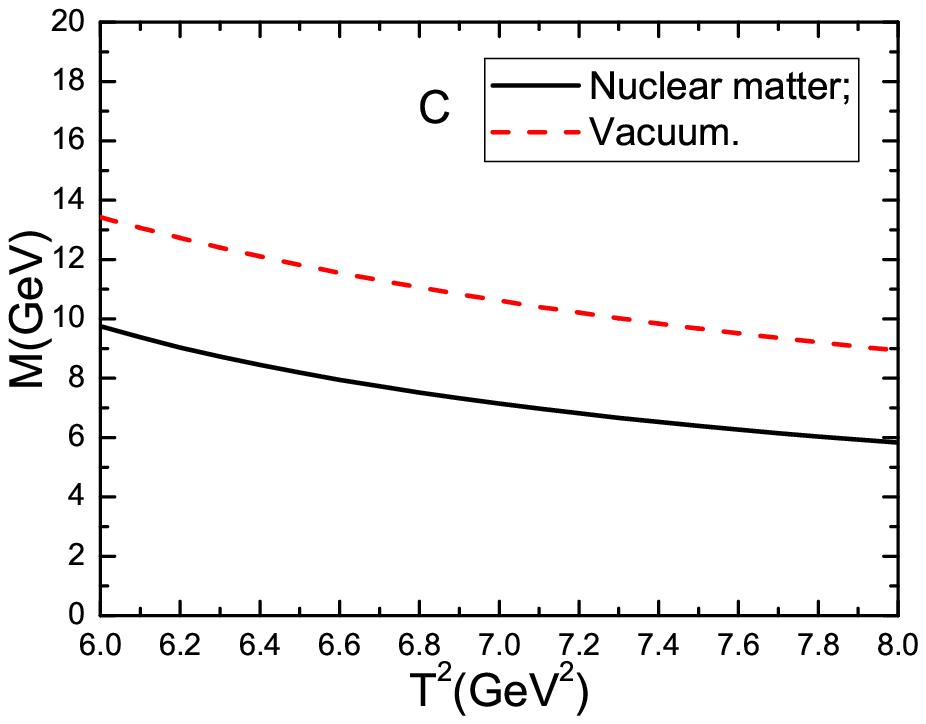}
 \includegraphics[totalheight=5cm,width=7cm]{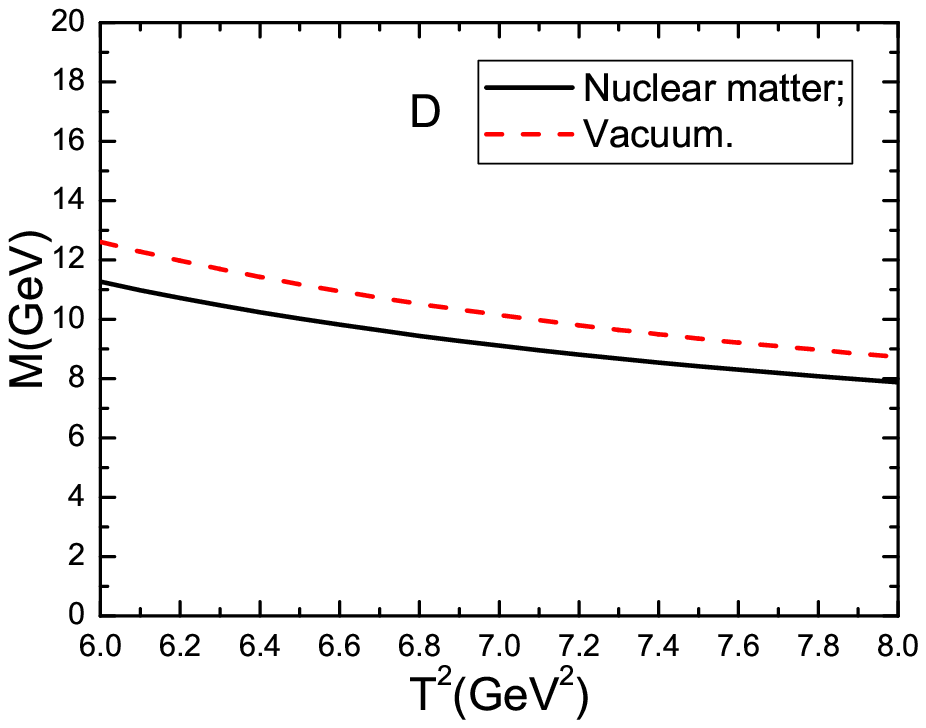}
 \caption{The  masses from the QCD sum rules in the vacuum and in the nuclear matter  versus the Borel parameter $T^2$, the $A$, $B$, $C$  and $D$ denote the
 $\Xi_{cc}$, $\Omega_{cc}$, $\Xi_{bb}$  and $\Omega_{bb}$ baryons, respectively. }
\end{figure}

\section{Conclusion}
In this article, we extend our previous works on the $\Lambda$-type and $\Sigma$-type heavy baryon states   to   study the
doubly heavy baryon states $\Xi_{QQ}$ and $\Omega_{QQ}$  in the nuclear matter using the QCD sum rules, and  derive three coupled QCD sum rules for
 the   masses, vector self-energies and pole residues  in the nuclear matter, then take the limit $\rho_N = 0 $ to recover the QCD sum rules in the vacuum, finally  obtain the values of the masses and pole residues   in the vacuum,   and the masses, vector self-energies and pole residues  in the nuclear matter.
  The numerical results indicate that the mass-shifts in the nuclear matter are about $\Delta M_{\Xi_{cc}}=-1.11\,\rm{GeV}$,
$\Delta M_{\Omega_{cc}}=-0.33\,\rm{GeV}$,
$\Delta M_{\Xi_{bb}}=-3.37\,\rm{GeV}$ and
$\Delta M_{\Omega_{bb}}=-1.05\,\rm{GeV}$, respectively, which can be confronted with the experimental data in the future.

\section*{Acknowledgments}
This  work is supported by National Natural Science Foundation,
Grant Number 11075053,  and the Fundamental
Research Funds for the Central Universities.

\end{document}